\documentstyle[12pt]{article}
\begin{document}

\title{From Fractal Cosmography to Fractal Cosmology}

\author{A. K. Mittal\thanks{\small Department of Physics, 
University of Allahabad, Allahabad - 211 002, India; 
mittal\_a@vsnl.com}, Daksh Lohiya\thanks{\small 
Department of Physics and Astrophysics, University of Delhi, 
New Delhi--7, India; dlohiya@ducos.ernet.in}}

\date{}
\maketitle 

\vspace{-1.2cm}
\begin{center}
\em Inter University Centre for  Astronomy and Astrophysics,
Postbag 4, Ganeshkhind, Pune 411 007, India
\end{center}




\begin{abstract}

	Assuming a fractal distribution of matter in the universe, 
consequences that follow from the General Theory of Relativity and the Copernican 
Principle for fractal cosmology
are examined. The change in perspective necessary to deal with a fractal universe
is highlighted. An ansatz that provides a concrete application of the {\bf Conditional  
Cosmological Principle} is provided. This fractal cosmology is obtained by arguments 
closely following those used in standard cosmology. The resulting model 
may play a significant role in the debate on whether the universe is a fractal or 
crosses over to homogeneity at some scale. This model may also be regarded as an idealized 
fractal model around which more realistic models may be built.
\end{abstract}

\pagebreak
\section*{} 
  
  $~~~~$Standard cosmology is based on the assumption of homogeneity 
 and isotropy of the Universe, the so-called Cosmological Principle, on
  scales greater than $10^8$ light years. During the last decade
  this assumption has come to be challenged. The number of galaxies $N(r)$ within
  a sphere of radius $r$, centred on any galaxy, is not proportional to
  $r^3$ as would be expected of a homogeneous distribution. Instead $N(r)$ is 
  found to be proportional to $r^D$, where $D$ is approximately equal to
  2. This is symptomatic of a distribution of fractal dimension $D$.

  	It has further been argued \cite{lab} that available evidence
  indicates that the fractal distribution of visible matter extends
  well upto the present observational limits without any evidence of 
  cross-over to homogeneity. This suggests that the entire Universe could
  be a fractal. At present this question is being hotly debated \cite{rees,piet}. 
  Controversy remains, interalia,
  for want of agreement on proper treatment of raw observational data
  sets.   As a result, the available observational evidence has not been
  able to pronounce unequivocally whether the Universe becomes homogeneous
  at large enough scales or whether it continues to remain a fractal 
  indefinitely.
  
  	Even though the evidence in favour of a fractal universe that
  does not homogenize on any scale is not undisputed, it must be remembered
  that no generally acceptable structure formation scenario has yet emerged within
  the framework of the standard big-bang cosmologies. 
  
  	In the absence of a viable alternative, it will be a useful exercise to
  take the fractal distribution as given and examine the consequences for a cosmological
  model that follow from the General Theory of Relativity and the Copernican Principle.
 
  	If the Universe does not homogenize on any length scale, can one have a 
  cosmological model consistent with this fractal picture,    
  which could also be reasonably concordant with observations.

  	Apart from controversies in interpretation of observational data, a
  major obstacle in the acceptance of a fractal cosmography is that no cosmolgical
  model has been proposed which would support this cosmography, be consistent
  with the General Theory of Relativity and satisfy the Copernican principle.
    
  	Assumptions, like the Cosmological Principle, helped give simplistic
  solutions to the Einstein's equations. When simplifying assumptions are
  dropped, chaotic solutions of Einstein's equations can be obtained \cite{hob}. 
  Chaotic dynamical systems are intimately linked with fractals. This 
  suggests  a need to explore the cosmological implications
  of chaotic solutions of Einstein's equations. In creating models of the 
  Universe, one invariably encounters fine-tuning problems. In order to
  overcome these problems, the inflationary scenario was proposed.
  However, this transfers the fine-tuning problem from cosmology to the
  underlying particle physics models. Thus there is no successful inflationary
  model, which can actually achieve the goals for which it was proposed.
  Models with fine-tuning problems are not regarded as acceptable because
  they lack predictive capability. In contrast to the approach followed 
  by cosmologists, it should be appreciated that chaotic dynamical systems,
  by definition, suffer from fine-tuning problems, yet they have been 
  useful in modeling physical phenomenon. This indicates that there is a
  need for seeking chaotic cosmological models exploiting the recent
  advances and successes in the fields of chaos and fractals. In particular
  one may expect that the fractal description, so successful in characterizing
  the statistics of dissipative eddies in turbulent flows, should also
  be useful for the statistical analysis of gravitational clustering.
  
  	Self-similarity is ubiquitous in nature. Perhaps it is the true 
  `Cosmological Principle'. Although, a fair amount of evidence has been
  collected in support of inhomogenous distribution of visible matter, there 
  is no evidence to contradict isotropy in a statistical sense from any
  galaxy. Mandelbrot \cite{mandl} proposed to replace the standard Cosmological
  Principle by the Conditional Cosmological Principle. According to this
  principle the Universe appears to be the same statistically from every 
  galaxy and in every direction. Thus the Conditional Cosmological Principle
  could allow one to obtain simplified solutions of the Einstein's equations,
  similar to those for the standard cosmology.
  
  	The conditional cosmological principle offers the hope that one can
  develop cosmological models along the lines of the standard model, while
  having a simple explanation of the fractal distribution of matter. Although
  the conditional cosmological principle was proposed about two decades ago,
  to the best of our knowledge, we are presenting the first ansatz that
  provides a concrete application of the conditional cosmological principle.

  	Homogeneity of the Universe in the standard model means that
  through each event in the Universe, there passes a spacelike 
  ``hypersurface of homogeneity''. At each event on such a hypersurface
  the density, pressure and curvature of spacetime must be the same.    
 
  	In a fractal universe, density is not defined at any point.
  The concept of density has to be replaced by that of a `mass measure'
  defined over sets. The mass measure as obtained by any observer
  will be the same. By an observer we will mean an observer moving 
  with the cosmological fluid. This precludes observers in a region
  of void. It should be noted that in a fractal, each of the points
  belonging to the fractal is on the same footing. But they are not
  on the same footing for the points not belonging to the fractal. 
  It is known that any sphere centred at a point not belonging to the
  fractal will be empty with probability 1.
  Because of this the conditional cosmological principle
  demands  an observer to be  situated
  on a galaxy (point of the fractal) and not in a region of void. 
    	 
 	We define a ``hypersurface of homogeneous fractality of
  dimension D'' as the hypersurface in which the mass measure over a
  sphere of radius R centred on the observer is proportional to $R^D$.
  We say that the Universe is a fractal universe of dimension D, if
  through each galaxy in the Universe, there passes a spacelike
  ``hypersurface of homogeneous fractality of dimension D''. 
  
  	Isotropy of the Universe means that, at any event, an observer
  who is ``moving with the cosmological fluid'' cannot distinguish any
  space direction from another by local physical measurements. 
  
  	It is widely believed that isotropy from all points of observation
  implies homogeneity. Thus an inhomogeneous Universe like a fractal could
  not be isotropic. It was shown from the observed isotropy of the Universe
  that the fractal dimension of the Universe could not differ appreciably from
  3; $(|D-3|< 0.001)$ \cite{peebls}. To counter this argument, 
  Mandelbrot \cite{chaosbook} demonstrated a 
  method of constructing fractals of any given dimension whose lacunarity
  could be tuned at will to make the distribution as close to isotropy
  (from any occupied point of the fractal) as desired. Thus in a fractal
  scenario isotropy from all galaxies may permit inhomogeneity to the extent
  of admitting homogeneous fractality. 
  
  	Isotropy of a fractal universe implies that the world lines
  of the cosmological fluid are orthogonal to each hypersurface of 
  homogeneous fractality. An observer moving with the cosmological 
  fluid can discover by physical measurements the hypersurface of
  homogeneous fractality relative to which the observer is at rest.
  His world line would be orthogonal to this hypersurface.
  
  	We now choose a hypersurface of homogeneous fractality $S_I$. To
  all the events on this hypersurface, one may assign coordinate time
  $t_I$. Galaxies on this hypersurface may be assigned coordinates $x^i$.
  We now let each galaxy evolve along with the cosmological fluid for 
  proper time $\tau$. We assign coordinate time $t= t_i + \tau$
  to the hypersurface formed by the galaxies. We let the spatial
 coordinates of each galaxy
  remain unchanged. This would correspond to a ``dust approximation'' for
  the fractal distribution. It is easy to see that this hypersurface will be
  a hypersurface of homogeneous fractality because each galaxy has the
  same environment and evolves under identical laws.
  
  	With coordinates to the galaxies being assigned in the above 
  manner, the interval between two galaxies with coordinates ($t, x^i$)
  and ($t + dt, x^i + dx^i$) will be given by 
$$
ds^2 = - dt^2 + g_{ij}(t,x)dx^idx^j \eqno{(1)}
$$
with isotropy demanding the time dependence to seprate as :
$g_{ij}(t,x) = a^2(t)\gamma_{ij}(x)$

	From isotropy and the Copernican Principle it is ordinarily shown \cite{mtw}
that the 3-metric $\gamma_{ij}(x)$ must yield the same curvature $K$ everywhere.
From this it follows that:
$$
\gamma_{ij}(x)dx^idx^j = d\chi^2 + \Sigma^2(d\theta^2 + \sin^2{\theta} d\varphi^2)
\eqno{(2)}
$$
where
$$
\Sigma \equiv \sin{\chi} \,\,\,\,\,\,\,\,\,\,\,\,\mbox{if $k 
\equiv {K\over{|K|}}= +1$ (positive spatial curvature)}\eqno{(3)}
$$

$$
\Sigma \equiv \chi  \,\,\,\,\,\,\mbox{if $k 
\equiv K = 0$ (zero spatial curvature)}\eqno{(4)}
$$

$$
\Sigma \equiv \sinh{\chi} \,\,\,\,\,\,\,\,\,\,\,\,\mbox{if $k 
\equiv {K\over {|K|}} = -1$ 
(negative spatial curvature)}\eqno{(5)}
$$

The FRW metric \cite{mtw} so obtained leads to a constant (same everywhere on the hypersurface of 
constant time) value of $G^{00}$ given
by 
$$
G^{00}_{\rm FRW}  =  3\left\{\left({{\dot a}\over a}\right)^2 + {k\over {a^2}}\right\}\eqno{(6)}
$$
However, this cannot satisfy Einstein's equations as $T^{00}$ is not constant 
everywhere for a fractal distribution of matter. If matter distribution on a constant 
time hypersurface satisfies the definition of homogeneous fractality given earlier, we 
know that
$$
\int_{S^3_P(R)}T^{oo}dV = M_P(R)\eqno{(7)}
$$
where $S^3_P(R)$ denotes a hypersphere of radius $R$ centered at $P$ on the hypersurface of constant
time $t$, and $M_P(R)$ is the mass enclosed in $S^3_P(R)$.
If the matter distribution on the hypersurface of constant time $t$ has a fractal dimension
$D$, we know that:

$$
M_P(R) =
\left \{
\begin{array}{lll}
&  C(t) R^D\\
& 0
\end{array}
\right.
\begin{array}{l}
\mbox{if P $\in$ the fractal }\\
\mbox{otherwise} 
\end{array} \eqno{(8)}
$$

Einstein's eqns. give
$$
\int_{S^3_P(R)}G^{oo}_{\rm fractal}dV = 8\pi M_P(R)\eqno{(9)}
$$
	As noted earlier, for a fractal distribution of matter, the concept of
density is undefined and has to be replaced by the notion of a measure on sets.
This implies that curvature (more precisely $G^{00}$) is not defined at any point.
However a measure may be defined to satisfy the integrated Einstein's equations
over any 3-volume in the constant time hypersurface. In the equation above
$G^{00}_{\rm fractal}$ may appear to be a function. It would be more appropriate to
regard it as an ansatz for defining a measure on sets containing the point P.
In the same way we could express the mass measure by using mass density $\rho$
proportional to $r^{D-3}$. Here $\rho$ would not mean the density at a point, but
merely an ansatz to compute the mass measure. In this way we can hope to express
the Einstein's equations for a fractal distribution of mass by a relation
connecting $G^{00}_{\rm fractal}$ to $\rho$, remembering clearly that these are not
functions but ansatz to compute measures. Thus the dependence of
$G^{00}_{\rm fractal}$ on $\chi$ and of $\rho$ on $r$ should not be seen as an indication
of inhomogeneity but rather as a means of concrete realization of how 
conditional cosmological principle satisfies the Copernican principle for all
occupied points of the fractal. 

Now our ansatz is essentially described by:

$$
G^{00}_{\rm fractal}(t,\chi,\theta,\varphi)  =
\left \{
\begin{array}{lll}
& f(\chi ) G^{00}_{\rm FRW}(t)  \\
& 0
\end{array}
\right.
\begin{array}{l}
\mbox{if P $\in$ the fractal }\\
\mbox{otherwise} 
\end{array} \eqno{(10)}
$$
On a constant time hypersurface, so long as the origin P is chosen as belonging to the 
fractal, the integrated $G^{00}$ measure on sets containing P will not depend on the choice 
of the origin. Thus, despite an apparent dependence of $G^{00}_{{\rm fractal}}$ on $\chi$,
the Conditional  Cosmological Principle will be satisfied.
Thus,
$$
\int_{S^3_P(R)}G^{00}_{\rm fractal}(t,\chi,\theta,\varphi)  dV = 
8\pi C(t)R^D\,\,\,\,\,\,\mbox{if P $\in$ fractal}\eqno{(11)}
$$
With $\chi$ chosen so that $R = \Sigma (\chi )a(t)$, eqns. (10) and (11) yield:
$$
4\pi G^{00}_{\rm FRW}(t) a^3(t)\int_0^\chi f(\chi )\Sigma^2(\chi ) \Sigma'(\chi )d\chi
= 8\pi C(t) a^D(t) \Sigma^D(\chi )\eqno{(12)}
$$
Hence:
$$
G^{00}_{\rm FRW}(t) = 2\nu C(t) a^{D-3}(t)\eqno{(13)}
$$
and 
$$
f(\chi )\Sigma^2(\chi ) \Sigma'(\chi ) = {D\over \nu}\Sigma^{D-1}(\chi ) \Sigma'(\chi )\eqno{(14)}
$$
or
$$
f(\chi ) = {D\over \nu}\Sigma^{D-3}(\chi )\eqno{(15)}
$$
where $\nu$ is a constant. We fix this constant by demanding $f(\chi) = 1$ for
$D = 3$. This gives $\nu = 3$. Thus,
$$
G^{00}_{\rm FRW}(t) = 6 C(t) a^{D-3}(t)\eqno{(16)}
$$
We denote $C(t)$ by $C_a$, because the scale factor is a function of time. 
When the scale factor is $a(t)$, the mass contained in a hypersphere of coordinate
radius $\chi$ is $C_a\{ a\Sigma (\chi )\}^D$. This mass remains unchanged 
 as the scale factor changes. Hence, 
$$
a^DC_a = a_o^DC_{a_o}\eqno{(17)}
$$
Therefore we simply get:
$$
3\left\{\left({{\dot a}\over a}\right)^2 + {k\over {a^2}}\right\} 
= 6 {a_o^D\over a^3} C_{a_o}\eqno{(18)}
$$

	There have been treatments of a fractal distribution 
of matter in the universe as a perturbation on a radiation dominated 
cosmology \cite{joyce}. In our approach, we have incorporated the fractal 
distribution of matter from the beginning. Eqn.(18) shows that the
dynamics of the scale factor due to a fractal distribution of matter 
is the same as in standard cosmology from an effective homogeneous 
density:
$$
\rho_{\rm eff} \equiv {3\over {4\pi}}{{a_o^DC_{a_o}}\over {a^3}} \eqno{(19)}
$$

	To summarize, instead of trying to explain the fractal clustering of the universe 
from hypothetical assumptions about the early universe, we have assumed as given, a 
fractal distribution of matter.  In the present context of ubiquity of chaos and fractals
in nature \cite{mandl} 
and observational evidence not ruling out  fractality, there is (at least) as much 
(rather much more of a) justification for making {\it a priori} assumption of 
homogeneous fractality,
as there  was for making the assumption of homogeneity at the time that the standard 
model was proposed.

	Just as the standard model follows naturally from the Cosmological Principle
when General Theory of Relativity is applied, the fractal model described in this article 
follows naturally from the conditional cosmological principle, once the necessary change in 
perspective to deal with fractal distributions is made.

	This model may play a significant role in the debate on whether the universe is a 
fractal or crosses over to homogeneity at some scale. It may be in order to recall
the words of Jim Peebles \cite{peebls}, ``The geometrical picture of a fractal
Universe is elegant, but since it has not been translated into a physical model
we can not discuss some of the precision cosmological tests''. We hope that the 
model described in this article would contribute to fill the gap and be regarded 
as an idealized model around which more realistic models of the universe may be built.

\vskip 1cm
\section*{Acknowledgements}   
   
We thank Inter University Centre for Astronomy and Astrophysics 
(IUCAA) for hospitality and facilities to carry out this research.
%\end{section}

\bibliography{plain}

\end{document}